\documentclass[apj]{emulateapj}
\usepackage{amsmath}

\newcommand{\Swift}{{\it Swift}}
\newcommand{\Suzaku}{{\it Suzaku}}
\newcommand{\Swifts}{{\it Swift }}

\newcommand{\ltsim}{\protect\raisebox{-0.5ex}{$\:\stackrel{\textstyle <}{\sim}\:$}}
\newcommand{\gtsim}{\protect\raisebox{-0.5ex}{$\:\stackrel{\textstyle >}{\sim}\:$}}

\slugcomment{Advances in Astronomy GRB special issue (accept)}
\shorttitle{A new era of sub-millimeter GRB afterglow follow-ups with the Greenland Telescope}
\shortauthors{Urata et al.}

\begin{document}

\title{A new era of sub-millimeter GRB afterglow follow-ups with the Greenland Telescope}

\author{
Yuji~\textsc{Urata}\altaffilmark{1,3}, 
Kuiyun~\textsc{Huang}\altaffilmark{2},
Keiichi~\textsc{Asada}\altaffilmark{3},
Hiroyuki~\textsc{Hirashita}\altaffilmark{3},
Makoto~\textsc{Inoue}\altaffilmark{3},
and
Paul T. P.~\textsc{Ho}\altaffilmark{3,4}
}

\altaffiltext{1}{Institute of Astronomy, National Central University, Chung-Li 32054, Taiwan, urata@astro.ncu.edu.tw}
\altaffiltext{2}{Department of Mathematics and Science, National Taiwan Normal University, Lin-kou District, New Taipei City 24449, Taiwan} 
\altaffiltext{3}{Academia Sinica Institute of Astronomy and Astrophysics, Taipei 106, Taiwan}
\altaffiltext{4}{Harvard-Smithsonian Center for Astrophysics, 60 Garden Street, Cambridge, MA 02138, USA}

\begin{abstract}
  A planned rapid submillimeter (submm) Gamma Ray Burst (GRBs)
  follow-up observations conducted using the Greenland Telescope (GLT)
  is presented. The GLT is a 12-m submm telescope to be located at the
  top of the Greenland ice sheet, where the high-altitude and dry
  weather porvides excellent conditions for observations at submm
  wavelengths. With its combination of wavelength window and rapid
  responding system, the GLT will explore new insights on GRBs.
  Summarizing the current achievements of submm GRB follow-ups, we
  identify the following three scientific goals regarding GRBs: (1)
  systematic detection of bright submm emissions originating from
  reverse shock (RS) in the early afterglow phase, (2)
  characterization of forward shock and RS emissions by capturing
  their peak flux and frequencies and performing continuous
  monitoring, and (3) detections of GRBs as a result of the explosion
  of first-generation stars result of GRBs at a high redshift through
  systematic rapid follow ups. The light curves and spectra calculated
  by available theoretical models clearly show that the GLT could play
  a crucial role in these studies.
\end{abstract} 


\section{Introduction}

Gamma-ray Bursts (GRBs) are among the most powerful explosions in the
universe and are observationally characterized according to intense
short flashes mainly in high-energy band (so-called ``prompt
emission''), and long-lasting afterglows observed in X-ray to radio
bands. Both types of radiation are sometimes extremely bright and can
be observed using small- and middle-aperture optical and near-infrared
telescopes \citep[e.g.,][]{z8,z8v2}. Because of their intense
luminosity, the highest-redshift($z$) in the reionization epoch
($z\gtsim8$) have been observed and have a high possibility for
discovery even at $z>10$ \citep{z8v2}. Although there are various
diversities (long/short duration of prompt $\gamma$-ray radiation, X-ray flares
associated with X-ray afterglows, and complex temporal evolutions of
afterglows), optical afterglow and host galaxy observations indicate
that the majority of long-duration GRBs occur as a result of the death
of massive stars \citep[e.g.,][]{stanek03}. Thus GRBs are unique and
powerful means of observing explosions first generation stars
(population III, POP-III). Understanding the diversity of the
astrophysical entities that cause GRBs is the subject of ongoing study
and represents one of the most prominent inquiries in modern
astrophysics.

Using GRBs to investigate the high-$z$ universe requires an understanding of
their radiation mechanisms. Confirming the existence of reverse shocks
(RS) and ascertaining their typical occurrence conditions are
critical. The GRB afterglow is believed to involve a relativistically
expanding fireball \citep[e.g.,][]{fireball}. The Interstellar Medium (ISM)
influences the fireball shell after it has accumulated, and
considerable energy is transferred from the shell to the ISM. The
energy transfer is caused by two shocks: a forward shock (FS) propagating
into the ISM, and an RS propagating into the shell.
Millimeter/Submillimeter (mm/submm) observations are the key element
in understanding the emission mechanism of GRB afterglows. They
provide ``clean'' measurements of source intensity and are unaffected
by scintillation and extinction. Hence, studies on submm properties of
the afterglow are likely to enrich the understanding of GRB physics.

In this paper, we review the status and achievements of submm
afterglow observations in \S2. In \S 3, we introduce the Greenland
Telescope (GLT) project and its advantages for GRB follow-up
observations. We also expect the GRB follow-ups in the era of GLT in
\S 4. On the basis of these advantages, we establish three science
goals by introducing expected light curves and spectra with the
expected sensitivity limit of GLT in \S 5. Finally, we summarize this paper in
\S 6.

\begin{figure*}
\epsscale{0.7}
\plotone{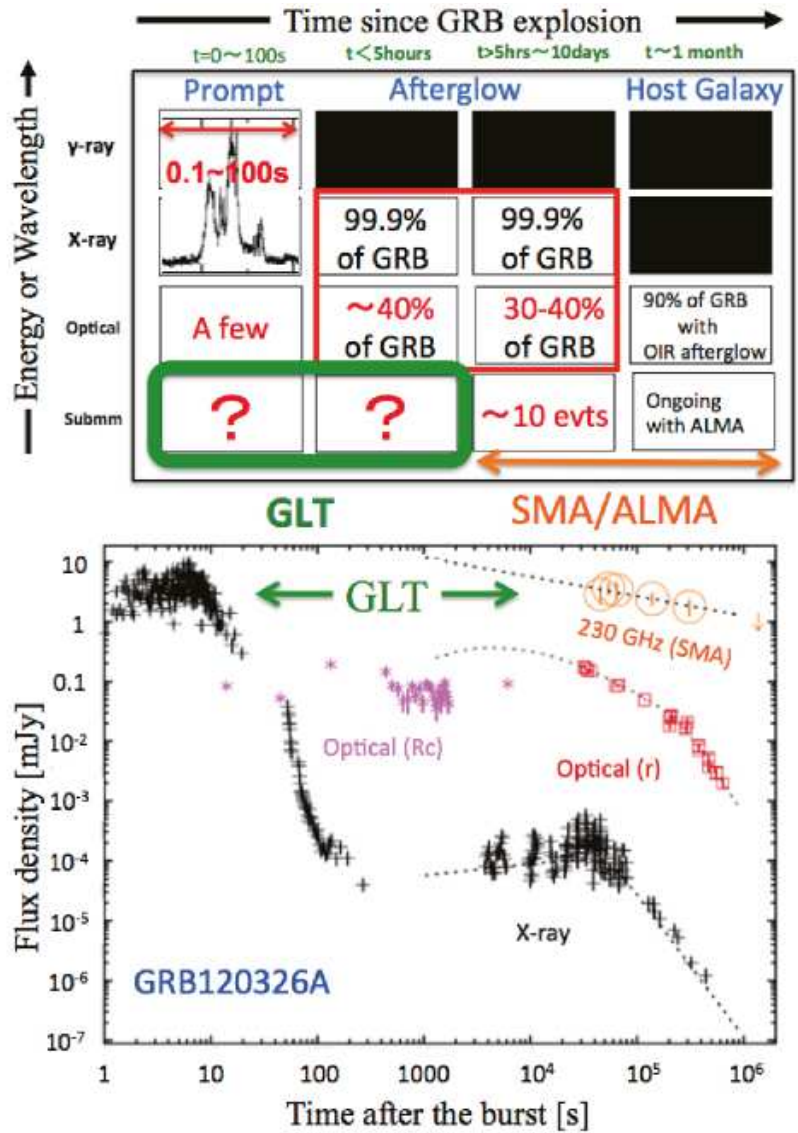}
\caption{(Top) Schematic summary of the GRB observational achievements along with time from the bursts in individual wavelength (from $\gamma$-ray to submm). (Bottom) One of actual light curves in X-ray, optical and submm with the earliest submm detection.} 
\label{fig:grbobs}
\end{figure*}

\section{Status and Achievements of Sub-millimeter Afterglow Follow-Ups}

\begin{figure}
\epsscale{1.0}
\plotone{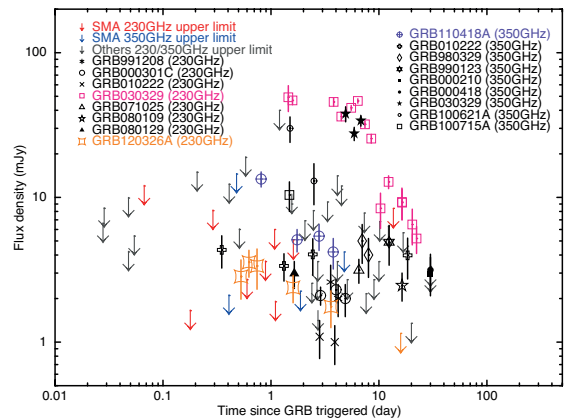}
\caption{Light curve summary of afterglow observations in submm bands (230 and 345 GHz). Red (230 GHz upper limits), blue (345 GHz upper limits), orange (GRB120326A), and purple (GRB100418A) points were obtained with the SMA. The SMA successfully monitored two of three well-observed submm afterglows, GRB100418A and GRB120326A.}
\label{fig:subobs}
\end{figure}

Numbers of dedeicated follow-up instruments of GRBs and afterglows
have been developed in $\gamma$-ray, X-ray \citep{swiftsat}, optical,
and near-infrared
\citep[e.g.][]{lt,rotse,raptor,piofsky,tarot,watcher,widget}.
Afterglow observations in X-ray and optical have been adequately
covered from very early phase including some fractions of $\gamma$-ray
prompt phase \citep[e.g.][]{990123,vestrand05,vestrand06,080319b}.
In addition, more than 300 afterglow observations have been made at
the cm wavelengths mainly using the Very Large Array
\citep[e.g.][]{chandra}.
However, submm has lagged behind X-ray and optical.
Fig. \ref{fig:grbobs} shows a schematic summary of achievements in GRB
observations according to wavelength and time range. It is obvious
that prompt afterglow observations are lacking at submm wavelengths.
The numbers of submm-detected events have been limited as summarized
in Fig. \ref{fig:grbobs} and \ref{fig:subobs}. Fig. \ref{fig:subobs}
shows all of afterglow observations in submm bands (230 and 345 GHz)
including upper limits. There have been only seven detections and
three well-monitored events (GRB030329, GRB100418A and GRB120326A) in
the submm bands. Unlike X-ray and optical observations, afterglow
monitoring in the submm band covers only the late phase of GRBs and
misses their brightening phases.
However, several of these observations, in conjunction with intensive
X-ray and optical monitoring, as in the GRB 120326A case
\citep{120326a}, have addressed crucial physical properties of
afterglow.  Hence, submm afterglow observations are crucial for
understanding the nature of GRBs.
In the following, we briefly summarize three well-monitored submm afterglow
cases.

{\bf GRB030329:} Because of the low redshift ($z=0.168$) origin and
bright optical afterglow ($\sim13$ mag at 0.1 days), numerous
follow-up observations were conducted at various wavelengths
\citep[e.g.][]{stanek03,smith03,tiengo03,klose04,kosugi04,urata04,tiengo04,kuno04,simon04,frail05,kohno05,smith05,gorosabel05,horst08}.
The 250-GHz monitoring follow ups were performed by the
MAMBO-2 bolometer array on the IRAM 30-m telescope \citep{030329v1}
and the IRAM Plateau de Bure Interferometer (PdBI) \citep{030329v2}.
The monitoring observation were conducted from 1.4 to 22.3 days after the
burst.  The light curve after $\sim$8 days exhibited a simple power-law
decay with a decay index of $-1.68$ \citep{030329v1}.
The value was consistent with that determined for the optical decay
after $\sim$0.5 days and incidated a common physical effect \citep{price03}.
These monitoring observations supported the
two-component jet model, in which a narrow-angle jet is responsible for
the high-energy emission and early optical afterglow; the radio
afterglow emission is powered by the wide-angle jet \citep{berger03,030329v1}.

{\bf GRB100418A:} The Sub-millimeter Array \citep[SMA;][]{sma} was
used to observe the submm afterglow from $\sim$16 hrs after the burst.
Continuous monitoring proceeded until 5 days after the burst, at which
point it became undetectable \citep{prealma}. As shown in
Fig. \ref{fig:subobs}, the submm light curve exhibited a significant
fading (equivalent decay power-law index of $\sim-1.3$) between 1 and
2 days and then exhibited a plateau phase until 4 days. The X-ray and
optical light curves showed a late bump peak at $\sim5\times10^{4}$ s
\citep{100418ax}.

\begin{figure*}
\epsscale{0.7}
\plotone{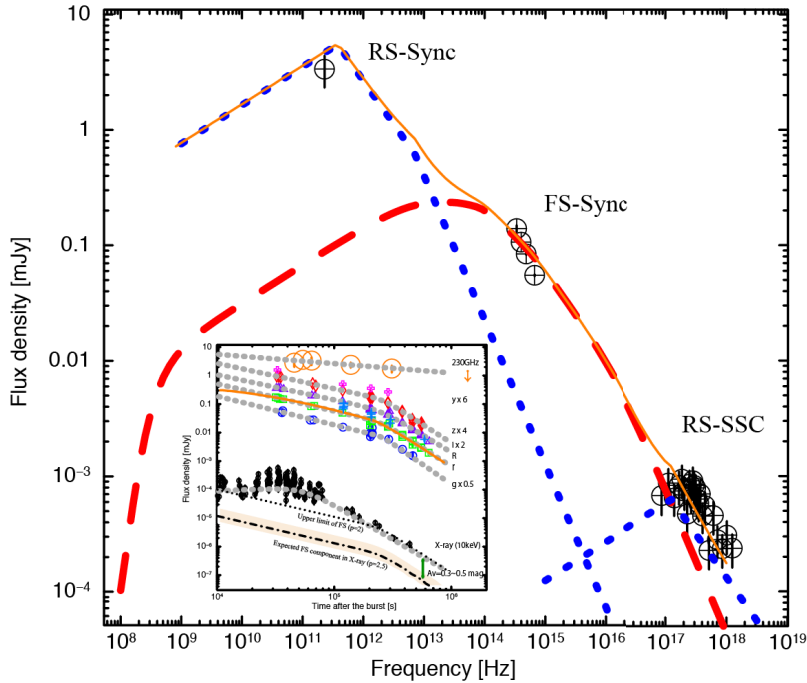}
\caption{Main panel: The spectral energy distribution of GRB120326A at $6.42\times10^{4}$ s after the burst \citep{120326a}. The red dashed line shows the forward shock synchrotron model spectrum calculated using the boxfit code \citep{boxfit} with the same parameters for the best modeling light curve shown in the sub panel. The blue dotted lines show the reverse shock synchrotron radiation and its self-inverse Compton component calculated based on \citet{ssc} using the observed values and model function for the forward shock component. Sub panel: X-ray, optical and submm light curves of the GRB120326A afterglow. The grey dotted lines show the best analytical fitted functions described in the text. The orange solid line shows the best modeling function for the $r$-band light curve obtained with the numerical simulation using boxfit.}
\label{fig:grbsed}
\end{figure*}

\begin{figure}
\epsscale{1.0}
\plotone{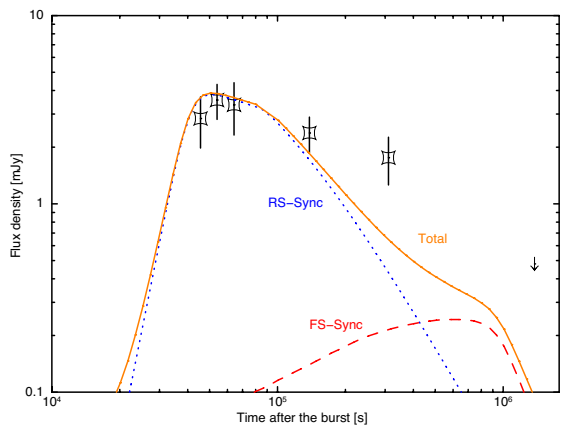}
\caption{Submm light curve of GRB120326A with RS and FS model functions.}
\label{grb120326a-rs}
\end{figure}

{\bf GRB120326A:} The SMA observation provided the fastest detection
to date among seven submm afterglows at 230 GHz
(Fig. \ref{fig:subobs}). In addition, comprehensive monitoring in the
X-ray and optical bands were also performed. These observations
revealed that the temporal evolution and spectral properties in the
optical bands were consistent with the standard FS synchrotron with
jet collimation ($6^{\circ}.7$). Furthermore, the X-ray and submm
behavior indicated different radiation processes from the optical
afterglow as shown in Fig. \ref{fig:grbsed}. Introducing synchrotron
self-inverse Compton radiation from RS is a possible solution that is
supported by the detection and slow decay of the afterglow in the
submm band.  As shown in Fig. \ref{grb120326a-rs}, the light curve
exhibited the slow temporal evolution ($\alpha_{submm}=-0.33$) between
$4\times10^{4}$ and $1\times10^{5} s$; this evolution is consistent
with the expected decay index of the RS with the $\nu_{obs} < \nu^{Sync}_{m,RS}$ condition \citep{120326a}.  Here, $\nu_{obs}$ and
$\nu^{Sync}_{m,RS}$ are the observing frequency and the characteristic
synchrotron frequency of the RS. and Because half of the events
exhibit similar X-ray and optical properties
\citep[e.g.,][]{panai,050319,normal} as the current event , GRB
120326A constitutes a benchmark requiring additional rapid follow ups
conducted using submm instruments such as the SMA and the Atacama
Large Millimeter/submm Array (ALMA).

To enhance afterglow studies, submm monitoring observations from an
early phase are required. Although rapid follow-ups have been
performed using with the SMA, as in the GRB130427A case (begining 1.6
h after the burst), these follow ups have been still limited;
specifically, we have failed to detect the GRB130427A afterglow with a
insufficient sensitivity ($\sim$10 mJy) for detecting the RS component
\citep{130427a1,130427a2,130427a3} caused by poor weather conditions.
The onging SMA programs have also been suffering from the poor weather
conditions of the Mauna Kea site. The condition of the Mauna Kea site
for the SMA and the James Clerk Maxwell Telescope (JCMT) is inferior
for submm observations comparing with other sites such as the ALMA
site. Five out of 13 observations with SMA (averaged responding time
is 11.3 hr) were made under the poor or marginal weather conditions.
Hence, weather condition for submm bands is one of the crucial points
for the time critical observations.
As demonstrated through JCMT observations \citep{13233,13259,13519,13554,14281,15174}, rapid follow-ups can be
managed by using existing submm telescopes with suitable follow-up
programs and observation system. The typical delay time of JCMT is hrs
scale (average of 59 min with 6 GRB observations).　Therefore, the
succession of the JCMT rapid response system is also desired in the future.
In addition, the constructions of dedicated submm telescopes based on
these experiences at the better observing site is required to conduct
systematic rapid and dense continuous follow-ups (sometimes
coordinated with several submm telescopes at the different longitude
for covering light curve within a day).

\section{Greenland Telescope}

The GLT is a state-of-the-art 12-m submm telescope to be located in
the Summit Station in Greenland. The aims of the project are
establishing a new submm very long baseline interferometer (VLBI)
station for the first imaging of shadow of supermassive black holes in
M87 \citep{inoue14}, and exploring a new terahertz frequency window
\citep{hirashita15} and time-domain astronomy in submm (e.g. this
paper). The expected first light of the GLT will be made in 2017/18.

\subsection{Site of the Greenland Telescope}

The Summit Station is located on top of the Greenland ice cap, at
72.5$^{\circ}$ N and 38.5$^{\circ}$ W (north of the Arctic Circle) at
a 3,200-m altitude.  The temperature is extremely low, especially in
winter, when the temperature reaches between $-$40$^{\circ}$C and
$-$60 $^{\circ}$C (a lowest temperature of $-$72$^{\circ}$C has been
recorded). Because of the combination of low temperature and high
altitude, considerably low opacity is expected. In 2011, a tipping
radiometer at 225 GHz was deployed to Greenland to monitor the sky and
weather conditions at the Summit Station, and measured opacities from
October 2011 to March 2014.  The best and the most frequent opacities
at 225 GHz were 0.021 and 0.04, respectively \citep{inoue14}. The
weather conditions are compatible with those at the ALMA site and
significantly better than those of the Manna Kea site \citep{gltsite}.
These low opacity and weather conditions are the advantage of the GLT
in managing submm and time critical observations including GRB
follow-ups with higher sensitivity (or short exposure cycle).

\begin{figure}
\epsscale{1.0}
\plotone{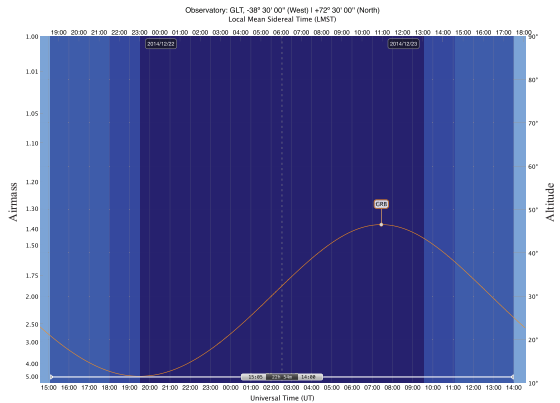}
\caption{A example of visibility plot in winter for the GRB that declination is 29 deg. Hatchs indicate the sun altitude lower than $-18$ (dark blue), $-12$ (blue), $-6$ (dark cyan), and 0 (cyan) deg, respectively.}
\label{fig:observer}
\end{figure}

\subsection{Planned instruments and expected sensitivities}

The GLT will be initially equipped with VLBI receivers at 86, 230, and
345 GHz. Whether large single-dish receivers as second-generation
instruments (e.g., submm hetrodyne arrays, a THz HEB array, and
bolometric spectrometer array) will be installed depends on the
scientific merits of the instruments and is still under
discussion. For GRB observations, three frequency bands of the VLBI
receivers are appropriate and therefore, the first-generation
receivers can be used for afterglow observations.  The current
expected sensitivities of VLBI receivers are 36 and 88 mJy with 1 s
integration at 230 and 345 GHz, respectively \citep{glt}. These are at
least two times better than those of SMA and JCMT.
We note that minimum integration time would be less than 0.5 s, since
duty cycle for positional switch will be 2 Hz. Longer integration will
be achieved as the accumulation of the short integrated data points.
Hence, the GLT with the receivers will provide sufficient
sensitivities to detect GRB counterparts at the 230 GHz band with
shorter exposures (e.g., 3$\sigma$ limit of 48.3 mJy with 5s, 10.8 mJy
with 100 s, 3.6 mJy with 15 min and 1.8 mJy with 1 h ). This shorter
exposure cycle is one of advanteges to charcterized temporal
evolutions of submm afterglows.
The field-of-view (FOV) of the receivers will be 25'' and 16'' at 230
and 345 GHz, respectively. These relatively narrow FOV require a
tiling observation for covering entire error region determined by
hard X-ray instruments (e.g. {\it Swift}/BAT) or accurate position
determination with X-ray afterglows.

We are also planning to install semi-automated responding system for the
GRB alerts to manage the rapid GRB follow-ups with secure procedures
at the extreme site. We enjoyed the prototype system at the Kiso
observatory \cite{urata04}. The pointing will be started using the
position determined by $\gamma$-ray instruments and additional
adjustment will be made responding to the position of X-ray afterglows.
By combination use of this system and site advantages including
visitility for targets as shown in Fig. \ref{fig:observer}, we will be
able to perform nearly real time follow-ups for GRBs whose declinations
are higher than 30 deg. A nearly video mode of continuous monitoring
will be managed in the first one day.

\section{Expected GRB follow-up observations in the GLT era}

To achieve successful observations, rapid follow ups using the GLT
will be coordinated through \Swifts and the planned Space-based
Multiband Astronomical Variable Objects Monitor
(SVOM)\citep{svom}. One of the obstacles to performing rapid follow
ups of {\it Swift}-detected GRBs is the poor visibility from ground-based
instruments. Although {\it Swift} has enabled detecting between 100 and 150
events per year, only $\sim10$ GRBs per year are observable from the
major astronomical observation sites (e.g., Mauna Kea, Chile) without
any delay from their $\gamma$-ray triggers, because of visibility
problems that occur when using ground-based instruments and the random
pointing strategy associated with {\it Swift} observations. According to
statistical data in 10 years of \Swifts observations, 10 to 12 events
per year could be observed from the early phase of the afterglow by
using currently existing telescopes within a $0-3$ hour delay by
maintaining a proper exposure time ($> 3-4$ h). The current fraction
of total \Swift/BAT pointing time to around antisun directions (sun
hour angle from 9 to 15 h) is limited to 30-40\%
\citep{swiftpoint}. The ideal location of the GLT will enable solving
this problem. As shown in Fig. \ref{fig:observer}, GRBs located at a
declination higher than 29 deg will always exhibit an elevation angle
higher than 12 deg over days. Hence, in winter, the GLT will be able
to begin rapid GRB follow ups without any delay caused by unfavorable
visibility and perform continuous submm monitoring over days. In
addition, these observation conditions are advantageous for observing
GRB afterglows that exhibit a rich diversity in various time ranges.
In summer, the 86 GHz receiver will be used when weather conditions
is unsuitable for observations at 230 and 345 GHz. 

{\it SVOM} (2020$\sim$) will be a timely mission for conducting rapid
GLT observations. The GRB detectors will observe antisun directions
that enable ground-based instruments to begin follow ups and
continuous long-term monitoring of markedly early cases immediately
 after receiving GRB alerts. The GRB detection rate of {\it SVOM} will
be $\sim$80 events per year, providing 10 to 20 events per year for
rapid GLT follow ups. In a typical GRB case, X-ray afterglows will be
observed immediately through a {\it SVOM} X-ray telescope (MXT)
\citep{mxt}, with the same strategy of X-ray observations of
\Swift. This provides a position accuracy of the counterpart of 2-3”,
which is sufficient to cover the position with the FOV of the GLT.

An additional crucial advantage of {\it SVOM} follow ups is the
capability of detecting of X-ray flashes (XRFs) and X-ray-rich GRBs
(XRRs) \citep{heteratio} and determining the prompt spectral peak
energy $E_{peak}$. Because of the slightly higher energy range of
\Swift/BAT (15-150 keV), sample collections of XRFs and related rapid
follow-ups have been entirely terminated. The $E_{peak}$ estimation of
the \Swift-detected GRBs (mainly XRRs and classical GRBs) has been
provided by joint spectrum fittings of \Swift/BAT and \Suzaku/WAM
\citep{wam}. Although spectral parameters of prompt emissions are
adequately constrained by these joint fittings
\citep[e.g.][]{starling08,hans,sugita,071010b,100414a}, the number of events is
limited.  This has caused a stagnation of the study of GRBs with
prompt characterization.
Because two of the prompt-emission-observing instruments onboard {\it
  SVOM}, ECLAIRs \citep{eclairs}, and the Gamma Ray Monitor (GRM)
\citep{grm} will cover the energy bands 4-150 keV and 30-5000 keV,
respectively, the $E_{peak}$ for most of {\it SVOM}-detected GRBs
would be determined. In addition, numerous XRRs and XRFs would be
detected with the $E_{peak}$ estimation through ECLIAS and
GRM. According to the {\it HETE-2} observations \citep{heteratio}, the
numbers of XRFs, XRRs, and GRBs were 16, 19, and 10, respectively. The
numbers of softer events (XRFs/XRRs) was considerably higher than that
of classical GRBs.
Because the lower-energy coverage of {\it HETE-2} (2-400 keV)
\citep{shirasaki} was similar to that of {\it SVOM}, a high volume of
XRF and XRR samples with $E_{peak}$ measurements can be
generated. This is likely to enhance the study of the origins of XRRs
and XRFs by enabling the determination of their physical parameters.
Hence, the GLT will be able to facilitate characterizing prompt and
late-phase submm afterglows of all three types of bursts for the first
time, providing crucial insights into the nature of XRFs and XRRs.

\section{Expected Science Cases}

On the basis of the summary of submm afterglow observations and the
GLT project, we established the following three scientific goals.

\subsection{Systematic detection of bright submm emissions}

It is believed that RSs generate short bright optical flashes
\citep[e.g.,][]{990123} and/or intense radio afterglows
\citep[e.g.,][]{990123radio}. According to the standard relativistic
fireball model, RSs are expected to radiate emissions in the long
wavelength bands (optical, infrared, and radio) by executing a
synchrotron process in a particularly early afterglow phase
\citep[e.g.,][]{RS}.

Detecting this brief RS emission and measuring its magnitude would
lead to constraints on several crucial parameters of the GRB ejecta,
such as its initial Lorentz factor and magnetization
\citep{zhang03}. Although RS emission has been detected in the optical
wavelength in several GRBs, early optical observations for most GRBs
have yielded no evidence of RS emission. The nondetection of RSs in 
optical bands could be an indicator of a magnetically dominated
outflow. Another possible reason for the nondetection is that the typical RS synchrotron
frequency is markedly below the optical band
\citep[e.g.,][]{melandri10}. Searching for RS emissions in the submm
wavelength would test these possibilities.
The comparison of early optical and submm temporal
evolution would enable studying the composition of the GRB ejecta.  If
an RS component were regularly detected in GRBs of which the early
optical light curve shows no evidence of RS emission, we would be able
to claim that GRB ejecta are baryonic in nature. The detection of RS
emissions in the submm band of most GRB would support the possibility
that GRBs are baryonic flow.

\begin{figure}
\epsscale{1.0}
\plotone{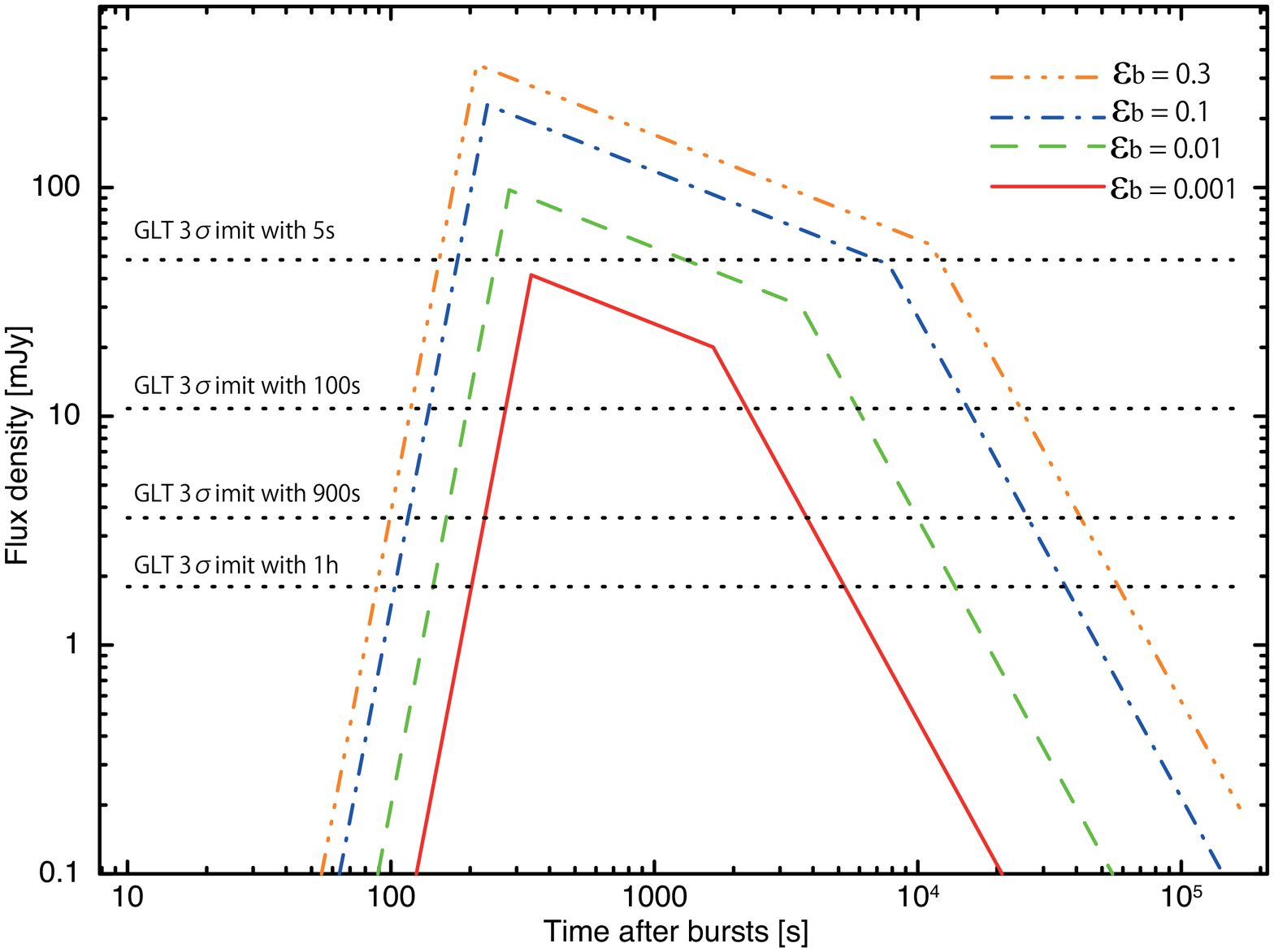}
\plotone{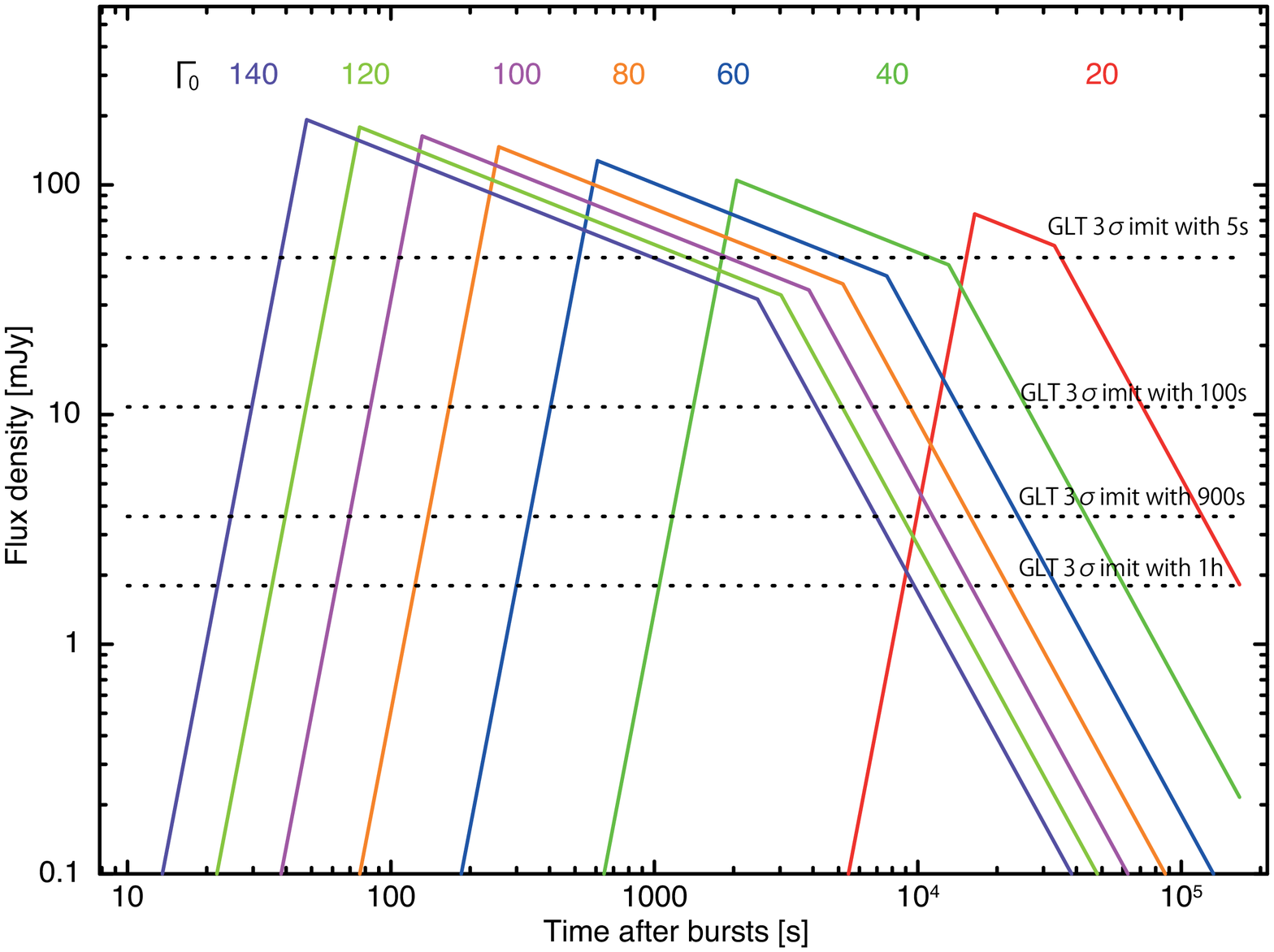}
\caption{(Top) Expected RS light curves in the 230GHz band with various $\epsilon_{B,RS}$ (0.001, 0.01, 0.1 and 0.3). Other physical parameters are fixed as $z=1$, $E=3\times10^{52}$, $\Gamma=80$, $n=1$, $\epsilon_{e}=0.3$ and $\epsilon_{B,FS}=0.01$. (Bottom) Expected RS light curves in the 230GHz band with various initial Lorentz factor $\Gamma_{0}$. Other physical parameters are $z=1$, $E=3\times10^{52}$, $n=1$, $\epsilon_{e}=0.3$, $\epsilon_{B,RS}=0.03$ and $\epsilon_{B,FS}=0.01$. The expected GLT 3$-\sigma$ limits with 900s and 1h exposures are indicated with dotted lines in both panals.}
\label{fig:rslc}
\end{figure}

One of the critical problems is that there has been no systematic
submm observational study in the early afterglow phase of GRBs.  As
shown in Fig.  \ref{fig:grbobs}, the number of events that have been
observed earlier than 1 day after bursts has remained 16 for some
time. The expected RS light curve for classical GRBs is fainter than 1
mJy at 1 day after bursts and therefore undetectable using currently
existing submm telescopes, expect for ALMA.
Fig. \ref{fig:rslc} shows the expected RS light curves based on
\citet{RS}, and \citet{ssc} with various magnetic energy densities of
RS $\epsilon_{B,RS}$ and an initial Lorentz factor $\Gamma_{0}$ in
comparison with the GLT sensitivity limit. In most of the cases shown in  Fig. \ref{fig:rslc}, the RS component faded away before
$1\times10^{5}$ s ($\sim1$ day). Hence, to detect and characterize RS
components, rapid ($\sim$ min scale) and continuous dense monitoring
within 1 day is required.
Although some of successful RS observations were made by SMA, CARMA,
VLA, and others with their open use framework
\citep[e.g.][]{120326a,130427a1,130427a2,130427a3}, dedicated radio
telescopes are strongly desired to make the systematic investigation.
In addition, dense monitoring with the same wavelength are required to
characterize the RS components. Because sparse monitoring even though
rapid detection is included failed to decode RS and FS components \citep[e.g.][]{failedvla}.
Therefore, the use of the GLT would initiate the era systematic the submm follow-ups.

In Fig. \ref{fig:rslc}, cases of low initial Lorentz factors (20, 40)
are provided showing XRRs and XRFs that are expected to be detected
using the planned GRB satellite SVOM.
The origin of the XRFs is not known. However, there are two major
models, namely (1) the failed GRBs or dirty fireball model
\citep[e.g.][]{failedgrb} and (2) the off-axis jet model
\citep{offaxis}.
According to the dirty fireball model, low-inital-Lorentz-factor
($\Gamma_{0} <<100$) GRBs produce a lower spectral peak energy
$E_{peak}$ in the prompt phase because of $E_{peak} \propto
\Gamma_{0}^{4}$ dependence, and it is therefore natural to attribute
this energy to XRRs and XRFs. The low Lorentz factors substantially
delay the RS peak (Fig. \ref{fig:rslc} bottom).
For the latter model, it is assumed that the viewing angle is considerably
larger than the collimation angle of the jet, and the high ratio of
X-ray to $\gamma$-ray fluence is caused by a relativistic beaming
factor when a GRB is observed through off-axis direction. The key
observable feature is the achromatic brightening optical afterglow
light curves, of which the peak time depends on the viewing angle \citep{nakar03,urata15}.
Hence, identifying a delayed RS peak through rapid GLT monitoring and
the prompt spectral characterization of {\it SVOM} will confirm and
identify the orgin of XRFs and XRRs.

\begin{figure}
\epsscale{1.0}
\plotone{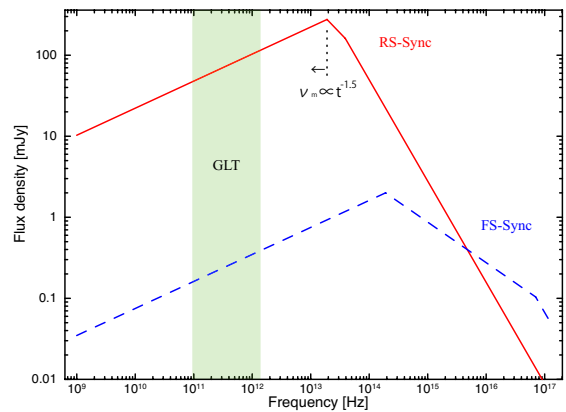}
\caption{Expected spectrum at the deceleration time with $z=1$, $E=1\times10^{52}$, $\Gamma=80$, $n=1$, $\epsilon_{e}=0.3$, $\epsilon_{B,RS}=0.03$, and $\epsilon_{B,FS}=0.01$. Synchrotron radiations from reverse and forward shock are indicated by red solid and blue dashed lines, respectively. A green hatched box indicates the frequency coverage of GLT.}
\label{fig:rsfs}
\end{figure}

\subsection{Characterization of forward and reverse shock emissions}

The spectral characteristics of FS and RS synchrotron emissions are
related as follows: $\nu_{m, FS}\sim
\mathcal{R}_{B}^{1/2}\mathcal{R}_{M}^{2}\nu_{m, RS}$, and $F_{max,
  RS}\sim \mathcal{R}_{M}\mathcal{R}_{B}^{-1/2}F_{max,FS}$
\citep[e.g.,][]{zhang03}, where $\mathcal{R}_{M}$
(=$\Gamma_{d}^{2}/\Gamma_{0}$) and $\mathcal{R}_{B}$
(=$\epsilon_{B,FS}/\epsilon_{B,RS}$) denote the mass and magnetization
ratio parameters, respectively. Here, $\Gamma_{d}$, $\epsilon_{B,FS}$,
and $\epsilon_{B,RS}$ are the Lorentz factor at the deceleration time,
the fractions of magnetic energy of RS, and FS, respectively.
RS emission is typically expected to be considerably brighter
($\sim100$ times) than FS emission as shown in
Fig. \ref{fig:rsfs}. Therefore, the emission from RSs is sensitive to
the properties of the fireball, and the broadband spectrum and light
curve evolutions of FS/RS can provide critical clues to understanding
GRBs.

Regarding FSs, afterglows can be described by synchrotron emissions
from a decelerating relativistic shell that collides with an external
medium. According to the FS synchrotron model, both the spectrum and
light curve consist of several power-law segments with related indices
\citep[e.g.,][]{sari,gao}. The broadband spectrum is characterized
according to the synchrotron peak frequency $\nu_{m, FS}$ and the peak
spectrum flux density $F_{max, FS}$. The peak spectrum flux is
expected to occur at low frequencies (from X-ray to radio) over time
(from minutes to several weeks) as $\nu_{m, FS}\propto t^{-3/2}$. The
peak spectrum flux density $F_{max, FS}$ is predicted to remain
constant in the cirucumburst model, whereas it decreases as $F_{max,
  FS} \propto t^{-1/2}$ in the wind model. Therefore, determining the
characterizing frequencies and peak fluxes by using temporal and
spectrum observations provides direct evidence of the FS/RS shock
synchrotron model and typical (or average) physical conditions of a
fireball.

\begin{figure}
\epsscale{1.0}
\plotone{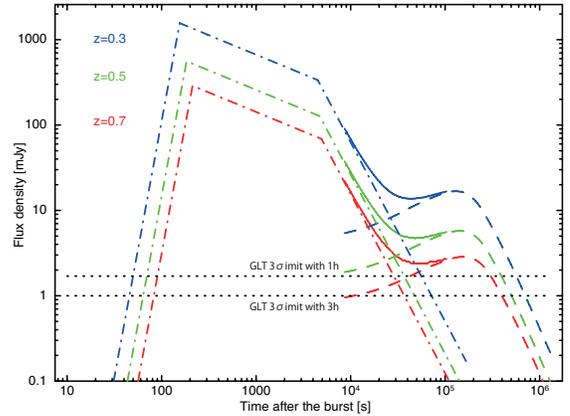}
\caption{Expected RS (dashed-dot) and FS (dashed) light curves in the 230GHz band at $z=0.3$, 0.5, and 0.7 with $E=3\times10^{52}$
erg, $n=$1 $cm^{-3}$, $p=2.1$, $\epsilon_{e}=0.3$, and $\epsilon_{B,FS}=0.01$. Solid lines indicate the total of RS and FS. GLT has also enough sensitivity to characterize the FS components for nearby ($z\ltsim0.7$) GRBs. The expected GLT 3-$\sigma$ limits with 1h and 3h exposures are indicated with dotted lines.}
\label{fig:fssync}
\end{figure}

Snapshots of the broadband spectrum and continuous monitoring of light
curves in the submm bands are essential to characterizing the
radiation of afterglow by decoding each
component. Fig. \ref{fig:fssync} shows the expected light curve in the
230 GHz band at $z=0.3$, 0.5, and 0.7.
We fix the rest of parameters as explosion energy $E=3\times10^{52}$
erg, circumburst number density $n=$1 $cm^{-3}$, the electron spectral
index $p=2.1$, the electron energy density $\epsilon_{e}=0.3$, and the
magnetic energy density of FS $\epsilon_{B,FS}=0.01$.
The brightening caused by the passing of the synchrotron peak
frequency $\nu_{m}$ can be detected to determine the FS component in
light curves.
The GLT has also enought sensitivities to detect FS component for nearby ($z\ltsim0.7$) events and the
monitoring determin their $\nu_{m,FS}$.
The expected $\nu_{m,FS}$ passing through time across the 230 GHz band is several
$\times10^{5}$ s (Fig. \ref{fig:fssync}). Therefore, the submm band is
suitable for decoding both RS and FS components.
Some of closure relations for FS and RS \citep{gao} will also
constrain components even if the light curve or spectrum observations
are sparsely performed.
For X-ray and OIR cases, the expected timescale is between several
minutes and $\sim2$ h after the burst. In this time range, detecting
the peak frequency directly is difficult because several additional
components (e.g., long-lasting prompt emission, X-ray flares etc)
characterize this phase.  For the MIR case, observations fully rely on
the satellite-based instruments. In this case, timely follow ups are
difficult because of operation confinement and limitation of number of
resources. Furthermore, the slow temporal evolution in the radio band
makes difficult to obtain simultaneous optical and X-ray
segments. This creates uncertainty regarding whether we observe the
same synchrotron components or not.
Hence, the GLT will provide unique results for nearby events
($z\ltsim0.7$) by facilitating the continuous monitoring.

Optical monitoring combined with the GLT will be required. As we describe above, RS components will be
caught by submm observations. For characterizing FS components,
multicolor optical monitoring is suitable because the temporal
evolution and spectrum of optical afterglow around 1 day after bursts
are well consistent with the expectation of the FS model.
Fig. \ref{fig:grbsed} shows one of the most appropriate examples of
the earliest submm afterglow detection procedures performed using the
SMA and related optical monitoring \citep{120326a}. Because of the
rapid submm monitoring, the FS and RS components were separated.
Conducting multifrequency monitoring by using the GLT requires rapid
optical follow ups, which will be conducted using our own optical
network EAFON \citep{lot, lee, eafon} and other ground-based optical
telescopes, as numerous observations have been conducted.

\subsection{Discovering of first-star explosions by using GRBs}

The findings regarding a high-redshift GRB at $z=8.2$ \citep{z8}
indicated the possibility of using GRBs to probe the processes and
environments of star formation as far back in time as the earliest
cosmic epochs.
Numerous theoretical models
\citep[e.g.,][]{pop3v1,pop3v2,pop3v3} show that some POP-III stars generate
GRBs as an end product. Hence, detecting the
signals of high-z GRBs has been a recent prominent objective in modern
astrophysics.

\begin{figure}
\epsscale{1.0}
\plotone{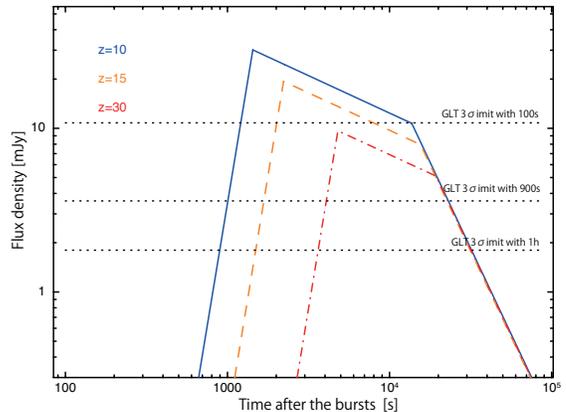}
\caption{Expected light curves of GRB afterglows at $z=$10, 15 and 30 at 230 GHz. Other physical parameters are fixed as $E=1\times10^{53}$, $\Gamma=80$, $n=1$, $\epsilon_{e}=0.3$, $\epsilon_{B,RS}=0.03$, and $\epsilon_{B,FS}=0.01$. The expected GLT 3$-\sigma$ limits with 900s and 1h exposures are indicated with dotted lines. GLT with rapid follow ups has sufficient potential to detect these higher-$z$ events.}
\label{fig:grbhighz}
\end{figure}

One of the prospective methods of identifying POP-III GRBs is to detect RS components in the
submm bands. Unlike OIR observations, submm observations provide clean
measurments of the source intensity, unaffected by extinction. Because
of the intense luminosity of the RSs, it is expected that the
radiation from high-z GRBs ($z>10-30$) can be observed if the
GLT is used with the rapid response system.
\citet{inoue} showed the RS component of POP-III GRBs at $z=15$ and 30 in the 300 GHz
band is substantially brighter than 1 mJy, and these bright RS
components will be detectable by using the GLT.
In addition, we simulated the expected RS light curves at $z=10$, 15,
and 30 based on \citet{RS} and \citet{ssc}. For this calculation, we assumed
$E=1\times10^{53}$ because the progenitor stars might be considerably
more massive than nearby events \citep[e.g.,][]{pop3v1}. Other
physical parameters are fixed as $\Gamma=80$, $n=1$,
$\epsilon_{e}=0.3$, $\epsilon_{B,RS}=0.03$, and
$\epsilon_{B,FS}=0.01$.
As shown in Fig. \ref{fig:grbhighz}, the GLT has great potential to
detect the high-$z$ GRBs even those at $z$=30 with the
rapid responding, shorter exposure cycle, and continuous dense
monitoring.
These initial detections of the GLT in the early phase will provide
the opportunity to conduct additional follow-ups using 30-m-class
telescopes such as the TMT.  Because these large telescopes typically
enable conducting follow ups for a limited number of events, the
target selections of the GLT observations will be critical.

The event rate of high-$z$ GRBs, which may be connected to the
star-formation rate in the early universe, is not known.
\citet{highzrate} estimated that the event rate of high-$z$ ($z>$7)
GRBs might be $\sim10$ events per year per steradian on the basis of
limited $\sim$100 \Swift-detected long GRBs with known redshift and
measured peak flux.
Their estimation showed that \Swift/BAT exhibited substantially high
redshift fractions ($\sim3.4\%$ at z$>$7), whereas \Swifts and related
follow ups detected only a few $z>7$ events.  Hence, uncertainty exist regarding the number of higher-$z$ events that
{\it Swift} has detected; thus, the frequency of such events is not
thoroughly understood because appopriate follow-ups in long wavelength
(e.g., IR, submm) have not been conducted. Actually,
$z\sim9$ GRB candidate \citep{z9} was also detected through \Swift.
Therefore, a continuous effort is required in this field of research,
despite a success rate that is typically low.
In addition, \citet{highzrate} expected that
{\it SVOM} will detect 0.1-7 events per year at $z>10$.
To detect these events, rapid follow-up coordination with
submm instruments will be crucial, because it is
impossible to identify higher-$z$ candidates within hours from bursts
with limited observational information.
Therefore, the installing a rapid responding system in the GLT will
enable us to perform high-$z$ GRB observations.

\section{Summary}

We briefly summarized the current achievements of submm follow-up
observations of GRBs. Although submm afterglow observations are
critical to understanding the nature of GRBs (e.g., GRB030329 and
GRB120326A), the number of successful observations has been
limited. This is because lack of dedicated submm telescopes has made
it difficult to perform rapid follow-up.

Furthermore, we introduced the single dish mode of GLT. The
development is ongoing and the expected first light of the GLT will be
made in 2017/18. The first light instrument will be the VLBI receivers
at 86, 230, and 345 GHz. The expected sensitivities of the receivers
are 36 and 88 mJy in 1 s integration at 230 and 345 GHz,
respectively. The GLT enables rapid and continuous submm monitoring of
GRB submm counterparts in the prompt phase.

According to the aforementioned situations and expected capabilities
of the GLT, we established the following three key scientific goals
regarding GRB studies; (1) systematic
detection of bright submm emissions originating from RS in the early
afterglow phase by conducting rapid follow ups, (2) characterization
of FS and RS emissions by capturing their peak flux and frequencies
through continuous monitoring, and (3) detection of the first star
explosions as a result of GRBs at a high redshift through systematic
rapid follow-ups.

Detecting RS emissions and monitoring light curves in the submm band
could lead to constraints on several crucial parameters of the GRB
ejecta, such as the initial Lorentz factor and magnetization. We
calculated the expected RS light curves by using various initial
Lorentz factors $\Gamma_{0}$ and the magnetic energy densities of RS
$\epsilon_{B,RS}$, and showed that these light curves could be
characterized through rapid follow ups of the GLT. Determining the
origins of XRRs and XRFs will also be a major focus of the GLT
together with the {\it SVOM} mission.

In addition to characterizing RS components, the GLT will be able to detect 
FS components of nearby ($z\ltsim$0.7) GRBs. Because the spectral
characteristics of the FS and RS synchrotron emissions are related,
characterizing both FS and RS components provides critical insights into
GRBs. We generated an expected submm light curve at
$z=0.3$, 0.5, and 0.7 and showed that the GLT can separate
RS and FS components through long-term and continuous monitoring, because
RS and FS components dominate earlier than $\sim4\times10^{4}$ s
and later than $\sim10^{5}$ s, respectively.

These two RS science topics will enhance the GRB studies as the probe
of a high-$z$ universe. Because of the existence of RS and its extreme
luminosity, it is expected that the radiation from high-z GRB
($z>10-30$) can be observed if the GLT is used with a rapid (hour
scale) response system. We simulated the expected RS light curves at
$z=10$, 15, and 30. These light curves showed that the GLT has
sufficient sensitivity to detect and characterize these events.  The
future {\it SVOM} mission may provide the capability to detect GRBs at
$z>10$; the establishment of close coordination using
longer-wavelength (e.g., infrared, submm) instruments will be crucial.
Because the rapid identification of counterparts at long wavelengths
is crucial for conducting additional further follow ups using
30-m-class telescopes such as the TMT will be used. Therefore, the GLT
could play a crucial role in detecting high-$z$ GRBs.

\acknowledgments 
The Greenland Telescope (GLT) Project is a collaborative project
between Academia Sinica Institute of Astronomy and Astrophysics,
Smithsonian Astrophysical Observatory, MIT Haystack Observatory, and
National Radio Astronomy Observatory.
We thank Shiho Kobayashi for useful comments. We also thank all
members of the GLT single dish science team organized at ASIAA.
This work is partly supported by the Ministry of Science and
Technology of Taiwan grants MOST 103-2112-M-008-021- (YU),
103-2112-M-001-038-MY2 (KA), 102-2119-M-001-006-MY3 (HH).

\end{document}